# Microstructure, tensile behavior and cyclic bendability of directly extruded biodegradable Zn, Zn-Mg and Zn-Mg-Sr wires

**L. Hlodák[a], K. Tesař[a], M. Lebeda[b,c], J. Duchoň[b], J. Kubásek[d], J. Čech[a], A. Školáková[b], J. Pinc[b,*]**

[a] *Department of Materials, Faculty of Nuclear Sciences and Physical Engineering, Czech Technical University in Prague, Trojanova 13, Prague, 120 00, Czech Republic*
[b] *FZU - Institute of Physics of Czech Academy of Sciences, Na Slovance 1999/2, Prague 8 182 21, Czech Republic*
[c] *Faculty of Mechanical Engineering, Czech Technical University in Prague, Technická 4, 16607 Prague 6, Czech Republic*
[d] *Institute of Metals and Corrosion Engineering, University of Chemistry and Technology, Technická 6, Prague 6, 166 28, Czech Republic*

Jan Pinc[*], Corresponding author, email: pinc@fzu.cz

Tel: +420 723 783 040

## Abstract

Zinc-based alloys are promising biodegradable implants, yet the trade-off between strength, ductility, and bendability in thin-wire form remains poorly understood. To address this, pure Zn (PZ), Zn-0.15Mg (ZM), and Zn-0.8Mg-0.2Sr (ZMS) wires of ~290 μm diameter were produced by direct extrusion, and their microstructure, texture, tensile, and bending behavior were characterized. $Mg_2Zn_{11}$ and $SrZn_{13}$ phases promoted particle-stimulated nucleation, refining the grain size from 35.8 μm in PZ to 6.8 and 3.3 μm in ZM and ZMS, respectively, weakening the basal-fiber texture, and raising the ultimate tensile strength at 37 °C from about 119 MPa for PZ to 291 and 334 MPa for ZM and ZMS, respectively. PZ deformed through twin transmission across weakly misoriented grains, producing serrated stress drops in tensile curves, with further straining accommodated by non-basal slip. ZM showed a single, aging-sensitive stress drop from localized twin nucleation and particle cracking, while ZMS showed distributed particle cracking without associated stress drops, consistent with its higher elongation to failure of 19 % at 37 °C. Mechanical properties remained broadly stable after 20 days of aging at 37 °C, with no measurable change in the nanoscale precipitates after 7 days. All compositions, however, showed markedly poor cyclic bendability accompanied by a twinning-detwinning mechanism directly observed by EBSD for the first time in zinc. These results establish single-step direct extrusion as a viable route to strong, fine-grained biodegradable Zn wires, while identifying poor bendability as the principal obstacle to bending-critical biomedical devices.

## 1 Introduction

Biodegradable metallic materials are a promising alternative to conventional permanent implants such as stainless steels, titanium alloys, and Co-Cr alloys, since they resorb after fulfilling their function, avoiding secondary removal surgery [1]. Magnesium, iron, and zinc are the three principal systems investigated to date [2], particularly suited to temporary fixation devices (sutures, staples, cerclage wires) where a permanent implant offers no long-term benefit once healing is complete [1]. Zn-based alloys, despite a favorable physiological role, are the most recently established of the three and remain comparatively less explored, especially in wire form.

Zinc is an essential trace element acting as a structural, catalytic, or regulatory component of several hundred enzymes and transcription factors [3], so moderate $Zn^{2+}$ release during degradation can largely be accommodated physiologically, provided the release rate stays controlled [3,4]. Bowen et al. [5] first convincingly demonstrated zinc as a feasible implant, showing that pure Zn wire implanted into rat arteries retained about 70 % of its cross-sectional area after four months while degrading safely. Although now established as one of the three principal classes of biodegradable metallic materials [4,6], no Zn-based implant has yet reached clinical application [2], mainly due to the low strength, limited ductility, and susceptibility to creep and natural aging of pure zinc [6].

The mechanical behavior of Zn is governed by its hexagonal close-packed structure, whose high c/a ratio of about 1.86 produces strongly anisotropic critical resolved shear stresses [7].

Basal slip activates at stresses roughly an order of magnitude lower than prismatic or pyramidal slip [8,9], so unfavorably oriented grains in textured material must instead twin or activate the harder non-basal systems [10,11], an interplay governing both tensile ductility and bending behavior [12].

Alloying combined with thermomechanical processing and severe plastic deformation refines microstructure and texture while preserving degradation behavior [13]. Magnesium is the most widely studied alloying element. It forms $Mg_2Zn_{11}$ intermetallic phases that promote grain refinement via particle-stimulated nucleation during dynamic recrystallization [14,15], raising strength through Hall-Petch strengthening. On the other hand, a decrease in ductility with increasing Mg content [16] and pronounced natural aging [17] were previously observed. Another biogenic element, strontium, promotes osteoblast proliferation and bone formation, and at low content forms fine $SrZn_{13}$ particles that refine the microstructure and improve corrosion passivation in ternary Zn-Mg-Sr alloys [18,19].

Most of this understanding comes from cast, rolled, or bulk-extruded material, whereas many biomedical applications require thin wires below 0.5 mm in diameter [20], which necessitates extrusion with optional drawing rather than simple scaling-down of bulk processing, thereby imposing high strains that strongly affect recrystallization and microstructure development [21,22]. Zn-Mg wires processed by hot extrusion and cold drawing have reached tensile strengths up to 270 MPa via $Mg_2Zn_{11}$-driven grain refinement [23,24], while more complex Li- and Cu-containing compositions relying on extensive cold drawing have reached the highest strengths yet reported for Zn wires [25,26]. Simpler, single-step directly extruded Zn-Mg and Zn-Mg-Sr thin wires remain underexplored. Nienaber et al. [27] studied pure Zn and dilute Zn-Mg wires, and Capek et al. [18] studied Zn-Mg-Sr only as a directly extruded bulk rod, while bendability remains considerably less studied than tensile behavior. Motivated by these considerations, the present work investigates single-step directly extruded pure Zn,

Zn-0.15Mg, and Zn-0.8Mg-0.2Sr wires. Their microstructure and texture are examined with attention to $Mg_2Zn_{11}$ and $SrZn_{13}$ particle-stimulated nucleation in grain refinement and texture weakening, together with their tensile deformation mechanisms and aging stability at 37 °C, and their bending behavior via EBSD and cyclic bending.

## 2 Materials and methods

### 2.1 Wire fabrication

Three alloys were investigated as directly extruded thin wires: commercially pure Zn (99.995 wt%), Zn-0.15Mg, and Zn-0.8Mg-0.2Sr (wt%), hereafter referred to as PZ, ZM, and ZMS. Prior to the extrusion, the alloys were prepared by melting and casting following the procedure described in [18], with composition confirmed by atomic absorption spectroscopy (AAS, Table S1, Supplementary Materials) and energy-dispersive spectroscopy (EDS). The microstructure of the annealed ZMS alloy was reported previously [28], with a similar microstructure expected for ZM based on comparable Zn-Mg alloys [29], since Sr additions mainly promote $SrZn_{13}$ particle formation and some grain refinement [30].

Cylindrical billets 7 mm in diameter and 22 mm in length were prepared from the homogenized ingots and from commercially pure Zn by electrical discharge machining (EDM), then machined to a final diameter of 6 mm and a length of 20 mm to remove surface contamination from the EDM brass wire. Before extrusion, all billets were pickled in 7 % Nital for 2 min and rinsed with ethanol. Direct hot extrusion was performed at 300 °C with a nominal extrusion ratio of approximately 1:400 and a ram speed of 0.2 mm/s, using GLEIT-µ HP 505 high-temperature lubricant to reduce friction. PZ, ZM, and ZMS wires with a final diameter of ~290 µm were successfully produced by this route. Due to the combined effects of exit-channel friction, elastic deformation of the die under loading, and thermal contraction of the wire upon cooling, the final wire diameter differed slightly from the nominal die-exit diameter corresponding to this ratio.

## 2.2 Microstructure, phase and texture characterization

Wires were embedded in Technovit 5000 (Kulzer Technik), ground, and polished using standard metallographic procedures, with a final colloidal silica step (OP-S, Struers) for ZM and ZMS and a water-free fumed silica suspension (Cloeren Technology GmbH) for PZ, since OP-S caused corrosion of PZ in the Cu-containing mounting resin. Longitudinal cross-sections relative to the extrusion direction were analyzed by field-emission-gun (FEG) scanning electron microscope (SEM, JSM-IT500HR, JEOL) at 20 kV, and phase area fractions were quantified from back-scattered electrons (BSE) micrographs by a custom MATLAB image analysis script (background flattening, thresholding, shape-based filtering) and visually validated. Crystallographic texture was analyzed by EBSD (Velocity EBSD camera, EDAX) at 30 kV with a 0.2–1.0 µm step size, indexed by spherical indexation and processed with TSL OIM Analysis 9. From EBSD, number fraction and average equivalent circle diameter were taken as the grain size $d$. Apart from grain size analysis, twins were excluded from their parent grain to avoid inflating the grain orientation spread (GOS). High-angle boundaries were defined by a misorientation angle of >15 °.

Lamellae were prepared perpendicular to the wire axis by focused ion beam lift-out (FEI Quanta 3D FEG DualBeam). Intermetallic phases and grain boundaries were analyzed by transmission electron microscopy (TEM) and scanning TEM (STEM) using an FEI Tecnai G2 F20 X-TWIN microscope (200 kV, double-tilt holder), with composition determined by STEM-EDS (30 mm$^2$ detector) and diffraction patterns evaluated using CrysTBox [31]. Precession electron diffraction (PED) was used to observe crystallographic orientation and phase maps at approximately 5 nm lateral resolution, analyzed via ACOM-TEM software (Nanomegas).

Volume distributions of secondary phases and pores in the as-extruded (AE) wires and after tensile testing were analyzed by µCT (Zeiss Xradia 610 Versa, 0.4 µm voxel size) and processed with Dragonfly (ORS, v. 2025.1) using histographic segmentation of the matrix, secondary phases, and pores, with a minimum object size of four voxels for reliable volume determination. For each identified particle, volume, surface area, aspect ratio, and Feret diameter were collected. Particle volume distribution as a function of distance from the wire surface was obtained from a signed distance map and normalized across shells, giving the volume concentration $C(r) = V_{\text{particles}}(r)/V_{\text{shell}}(r) \times 100$ %, as a function of distance $r$ from the wire axis based on calculated volumes $V$.

Small-angle X-ray scattering (SAXS) measurements were performed on a SAXSpoint 5.0 instrument (Anton Paar, Cu Kα, λ = 0.15418 nm, sample-to-detector distance 1621 mm), each collected for 30 min in vacuum on AE wires at RT and in situ at 37 °C after 0, 3, and 7 days of heat exposure to assess the thermal stability of nanoparticles in ZM and ZMS. Prior to measurements, the wires were ground on both sides to a rectangular cross-section, with a total sample thickness of up to 100 µm, to enable transmission within the sample. Two-dimensional patterns were azimuthally integrated into 1D profiles $I(q)$ (1000 radial bins), where $q$ denotes the scattering vector, with normalization by transmittance and background correction against a blank holder measurement. Profiles were fitted in SasView (v. 6.1.2) [32] with a polydisperse sphere model (lognormal size distribution) over $q \geq 0.02$ nm$^{-1}$.

### 2.3 Mechanical properties

Nanohardness ($H_{\text{IT}}$) and indentation modulus ($E_{\text{IT}}$) of the matrix and secondary phases were measured with an NHT$^2$ tester (Anton Paar, Berkovich tip), using a 0.5 mN maximum force with 15 s loading, 10 s hold, and 10 s unloading. These loading conditions were used to minimize the matrix influence on the measurement of the secondary phases and the effect of creep, which is significant for Zn even at RT [6]. Data were evaluated by the Oliver-Pharr method [33] according to ISO 14577.

Tensile tests were performed at RT and at 37 °C (using an Instron 3119-605 chamber) on an ElectroPuls E3000 system (Instron, 5 kN load cell, 10 mm gauge length, initial strain rate $10^{-3}$ $s^{-1}$), giving the 0.2 % offset yield strength ($R_p$), ultimate tensile strength ($R_m$), and elongation to failure ($A$) from the engineering stress-strain curves, calculated from the crosshead displacement. Fractography on samples tested at 37 °C was conducted using FEI Quanta 3D field-emission-gun DualBeam SEM using secondary electrons (SE, 10kV). Specimens aged at 37 °C for 5, 10, and 20 days were subsequently tested at room temperature (RT). All aging experiments in the present work were conducted in air. A representative specimen per alloy tested at 37 °C was selected for SEM and EBSD characterization at the neck and approximately 350 μm away, for both test temperatures. Because ZM showed unexpected tensile curve behavior, EBSD was additionally performed on a specimen for which the 37 °C test was manually interrupted before fracture.

A custom bending apparatus shown in Fig. S1 (Supplementary Materials) was used to compare the three compositions under three bending modes applied to separate specimens: a single bending (B1), a bending and straightening (S1), and a bending with subsequent bending in the opposite direction (B2), each followed by EBSD mapping of the bent region. Cyclic bending was performed with the same apparatus to determine the number of cycles to failure for each composition as a first approximation of fatigue-like behavior.

# 3 Results

## 3.1 Microstructure and phase analysis

Microstructure and texture of pure zinc (PZ), Zn-0.15Mg (ZM), and Zn-0.8Mg-0.2Sr (ZMS) wires analyzed by SEM and EBSD are shown in Fig. 1 on longitudinal cross-sections relative to the extrusion direction (ED). As-extruded samples showed no pores, cracking, or oxides visible by secondary electron (SE) or back-scattered electron (BSE) imaging (Figs. 1a–c). Needle-like features in PZ correspond to $\{10\bar{1}2\}$ compression twins (CTs), confirmed by inverse pole figure (IPF) maps with a misorientation angle of 93.5 ± 0.2 ° (close to the reported 94 ° for $\{10\bar{1}2\}$ twinning in zinc [34]), covering only a minor fraction of the PZ cross-section (Fig. 1d) and possibly originating from metallographic preparation rather than extrusion.

ZM and ZMS wires show two intermetallic phases using BSE in Figs. 1b, c. Dark $Mg_2Zn_{11}$ and, in ZMS, white $SrZn_{13}$, both confirmed by TEM in Fig. 3. In ZM, $Mg_2Zn_{11}$ has irregular globular-to-sharp-edged shapes with an area fraction of ~2.0 %, some elongated and fragmented along the ED into tail-like structures. In ZMS, $Mg_2Zn_{11}$ shows the same morphology but a higher area fraction of ~11.6 % and larger, more extensively distributed fragmented tails, while $SrZn_{13}$ shows no ED elongation, a smaller size, and a lower area fraction of ~1.1 % based on image analysis. Nanohardness $H_{IT}$ reached 1315 ± 179 MPa for the Zn matrix, 2945 ± 400 MPa for $Mg_2Zn_{11}$, and 8796 ± 244 MPa for $SrZn_{13}$ phases, with indentation modulus $E_{IT}$ equal to 90 ± 23, 117 ± 20, and 137 ± 5 GPa, respectively. Both phases were already observed in the as-cast, annealed ZMS alloy in our previous work as a continuous interdendritic $Mg_2Zn_{11}$ network with embedded submicron $SrZn_{13}$ phases [28], reflecting the low solubility of Mg and Sr in Zn [13]. After extrusion, $Mg_2Zn_{11}$ became finer and fragmented into ED-aligned rows, consistent with other extruded or drawn Zn-based alloys [18,25], while $SrZn_{13}$ is dispersed further relative to the annealed state, without elongation or fragmentation, likely reflecting its nearly three-fold higher hardness relative to $Mg_2Zn_{11}$.

IPF maps in Figs. 1d–f show unindexed regions where secondary-phase etching degraded pattern quality. Grain size (GS) decreased with alloying, from 35.8 ± 28.0 μm (PZ) to 6.8 ± 3.3 μm (ZM) and 3.3 ± 2.0 μm (ZMS). A mild center-to-edge difference in GS suggested by the IPF maps is equal to 6.3 ± 3.5 vs 5.8 ± 3.2 μm for ZM, and 3.4 ± 2.0 vs 3.2 ± 1.7 μm for ZMS, both within the scatter of the data and not statistically significant. Sharp interfaces spanning several grains along the ED were frequent in ZMS (Fig. 1f, black arrows), occasional in ZM (Fig. 1e), and completely absent in PZ (Fig. 1d), with further analysis in Figs. 3i–k. Pole figures (PFs) in Figs. 1g–i show a strong fiber texture, with (0001) planes perpendicular to the ED in all compositions, typical of extruded Zn alloys [16,35] and driven by basal <a> slip rotating the c-axis away from the extrusion axis during extrusion at 300 °C [36]. The multiples of a random distribution (m.r.d.), reflecting PF intensity, were the highest for PZ (17.1) and lowest for ZMS (3.9), with ZM slightly higher (4.4).

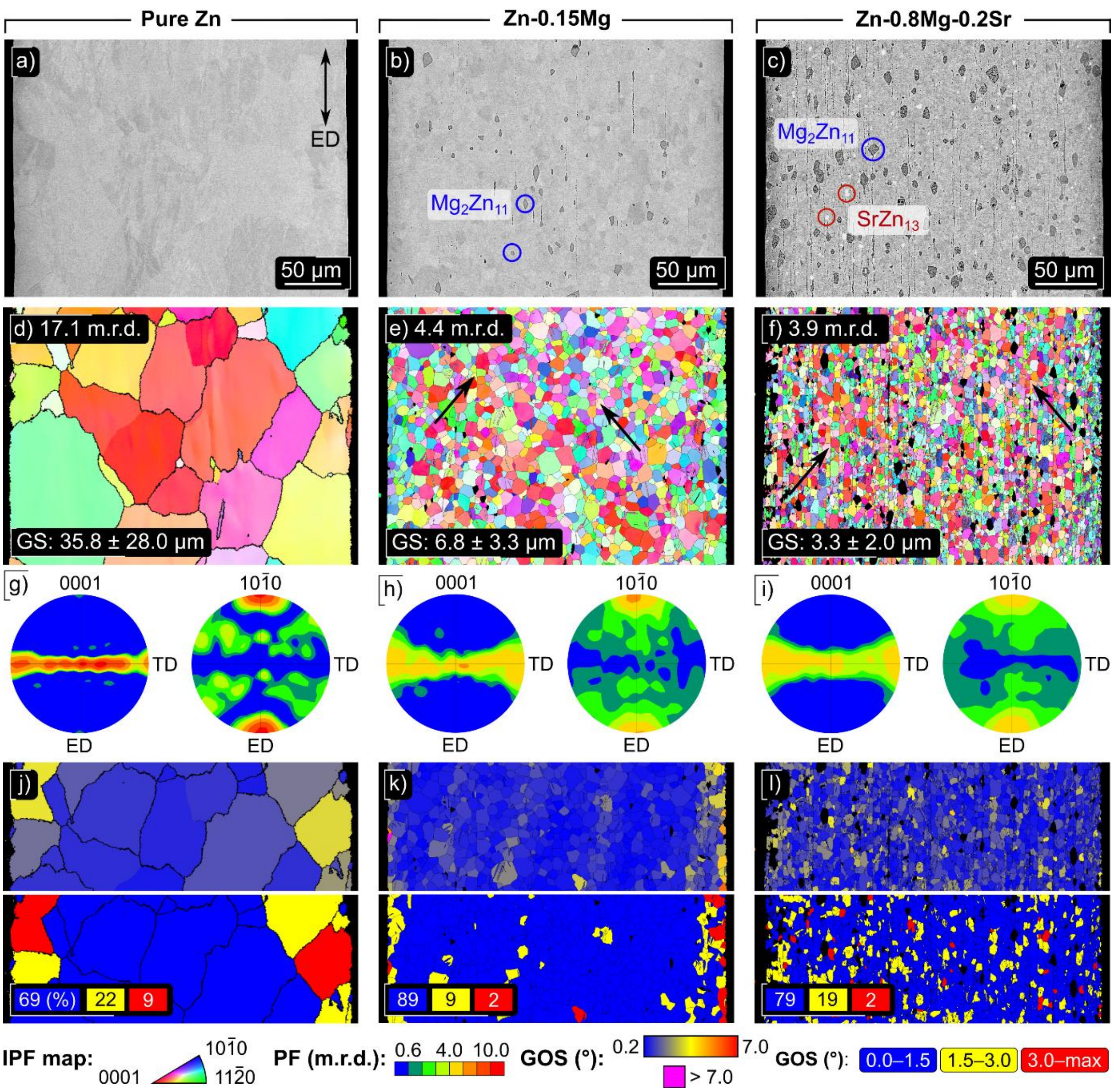

Fig. 1. a)–c) and d)–f) BSE SEM images and IPF maps of the as-extruded microstructures. g)–i) PFs of each wire composition. PFs, corresponding m.r.d. values, and grain size (GS) were calculated from a larger field of view than that shown in IPF maps. j)–l) Corresponding GOS maps for the IPF maps in two color-codings.

GOS maps (Figs. 1j–l), used to assess dynamic recrystallization (DRX), distinguish strain-free recrystallized grains (low GOS) from deformed or partially recovered ones (elevated GOS) [37]. Both continuous-scale and three-interval discretized GOS maps were used, providing a baseline for comparison with the behavior observed after subsequent tensile deformation and bending relative to the AE state. All compositions show a predominantly low-GOS, largely recrystallized microstructure. When GOS populations are quantified into recrystallized (≤1.5 °), DRX-deformed (1.5–3.0 °), and deformed (>3.0 °) fractions in Figs. 1j–l, PZ shows the lowest recrystallized fraction (69 %) and highest deformed fraction (9 %), with the remainder (22 %) being DRX-deformed. Alloying with Mg and Mg+Sr raised the recrystallized fraction to 89 % and 79 % and shifted the deformed fraction to 9 % and 19 %, respectively, while 2 % remained deformed for both alloys. These thresholds of 1.5–3.0 ° are slightly lower than the 2–5 ° typically applied to Mg alloys (for example, Hadadzadeh et al. [38]), but close to the 1.6 ° threshold used by Pan et al. [39] for a room-temperature-deformed Zn alloy, consistent with the low homologous recrystallization temperature and correspondingly rapid DRX kinetics of Zn-based alloys near RT. No change in GOS, GS, or texture was observed for any composition after aging at 37 °C for up to 20 days (for PZ, see Figs. S2a and b in Supplementary Materials).

X-ray micro-computed tomography (µCT) of the intermetallic particles (Figs. 2a–d) gave volume fractions (vol %) of 1.3 and 4.8 for ZM and ZMS. The mean particle radius and aspect ratio distributions (Figs. 2c and d) show ZM dominated by a relatively uniform population of fine particles with radii of 1–6 µm, while ZMS exhibits a broader distribution extending to larger radii up to 10 µm, consistent with a coarser eutectic structure in the as-cast ZMS alloy [28]. In contrast, BSE SEM analysis gave higher area fractions of 2.0 and 12.7 % (ZM vs ZMS) and resolved more elongation of phases in ZMS, with both discrepancies attributed to the voxel resolution of 0.4 µm in µCT (for further discussion, see Section 3, Supplementary

Materials). Furthermore, an almost perfectly circular cross-section of wire was observed via μCT, confirming good extrusion process reproducibility, with 3D and 2D reconstructions visible in Fig. S3 (Supplementary Materials).

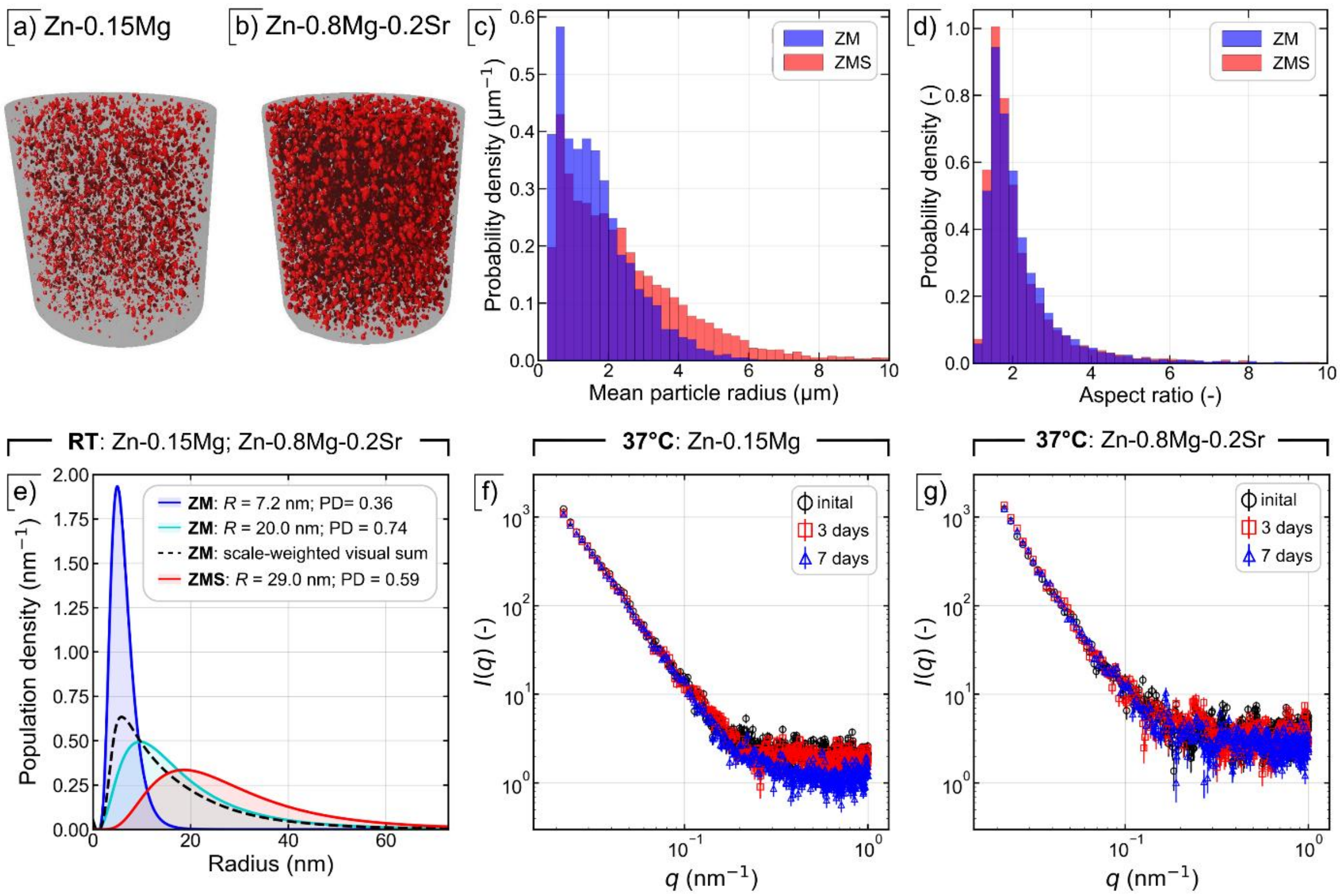

Fig. 2. a), b) 3D reconstructions of intermetallic particles with corresponding c) mean particle radius and d) aspect ratio. e) SAXS particle radius distributions of the alloyed wires at RT. f), g) One-dimensional SAXS scattering profiles measured at 37 °C and after exposure to 37 °C for 3 and 7 days.

SAXS, sensitive to precipitates below the resolution of μCT, was used to characterize nanoscale precipitates in Figs. 2e–g. Fitted SAXS curves measured at RT are shown in Fig. S4 (Supplementary Materials). The ZM profile ($q \geq 0.02$ nm$^{-1}$) required two independent polydisperse-sphere populations, each with a lognormal radius distribution and polydispersity PD = $\sigma_R/\bar{R}$ ($\sigma_R$ standard deviation, $\bar{R}$ mean radius). These comprised a larger population of mean radius ~20.0 nm (PD = 0.74) and a minor population of 7.2 nm (PD = 0.36) (Fig. 2e). The larger population dominates the SAXS response, being weighted toward larger particle volumes, while the smaller one is needed to describe the high-$q$ region. The fit therefore represents an intensity-weight rather than a number distribution. For ZMS, a single population sufficed, giving a broader distribution with mean radius ~29.0 nm (PD = 0.59), indicating one characteristic length scale rather than the two separated populations seen for ZM. Neither composition showed measurable change in SAXS response after 7 days at 37 °C in Figs. 2f and g. ZM showed only a slight, near-background intensity decrease in the high-q region (0.1–1.0 $nm^{-1}$) with exposure time, not reliably attributable to a structural change, while ZMS showed no visible change. The nanoparticles in both alloys thus appear stable under physiological-temperature conditions over the investigated period.

Higher-magnification BSE images of the ZM and ZMS wire microstructure show the $Mg_2Zn_{11}$ and $SrZn_{13}$ phases in Figs. 3a and b. Both types of phases were selected for TEM analysis. Bright-field (BF) TEM with the corresponding selected-area electron diffraction (SAED) pattern confirmed the $Mg_2Zn_{11}$ phase, indexed along the $[12\bar{3}]$ zone axis (Figs. 3c and d), while fast Fourier transform (FFT) of high-resolution TEM (HRTEM) confirmed the $SrZn_{13}$ phase, indexed along the $[2\bar{3}\bar{1}]$ zone axis (Figs. 3e, f). An ASTAR orientation map and corresponding phase map acquired by PED from a ZMS region containing both particles (Figs. 3g and h) showed that neither phase forms a single continuous domain in a plane perpendicular to the ED. Both $Mg_2Zn_{11}$ (green) and $SrZn_{13}$ (blue) are interspersed with numerous small regions of the surrounding Zn matrix (red), giving each particle a subdivided, mosaic-like appearance, with the $SrZn_{13}$ region itself composed of several sub-domains of distinguishable crystallographic orientation. The fragmented appearance of both phases resembles the dislocation-assisted fragmentation reported for eutectic $Mg_{17}Al_{12}$ in Mg-Al alloys reported by Zhao et al. [40], though confined to a single compact region rather than dispersed broadly.

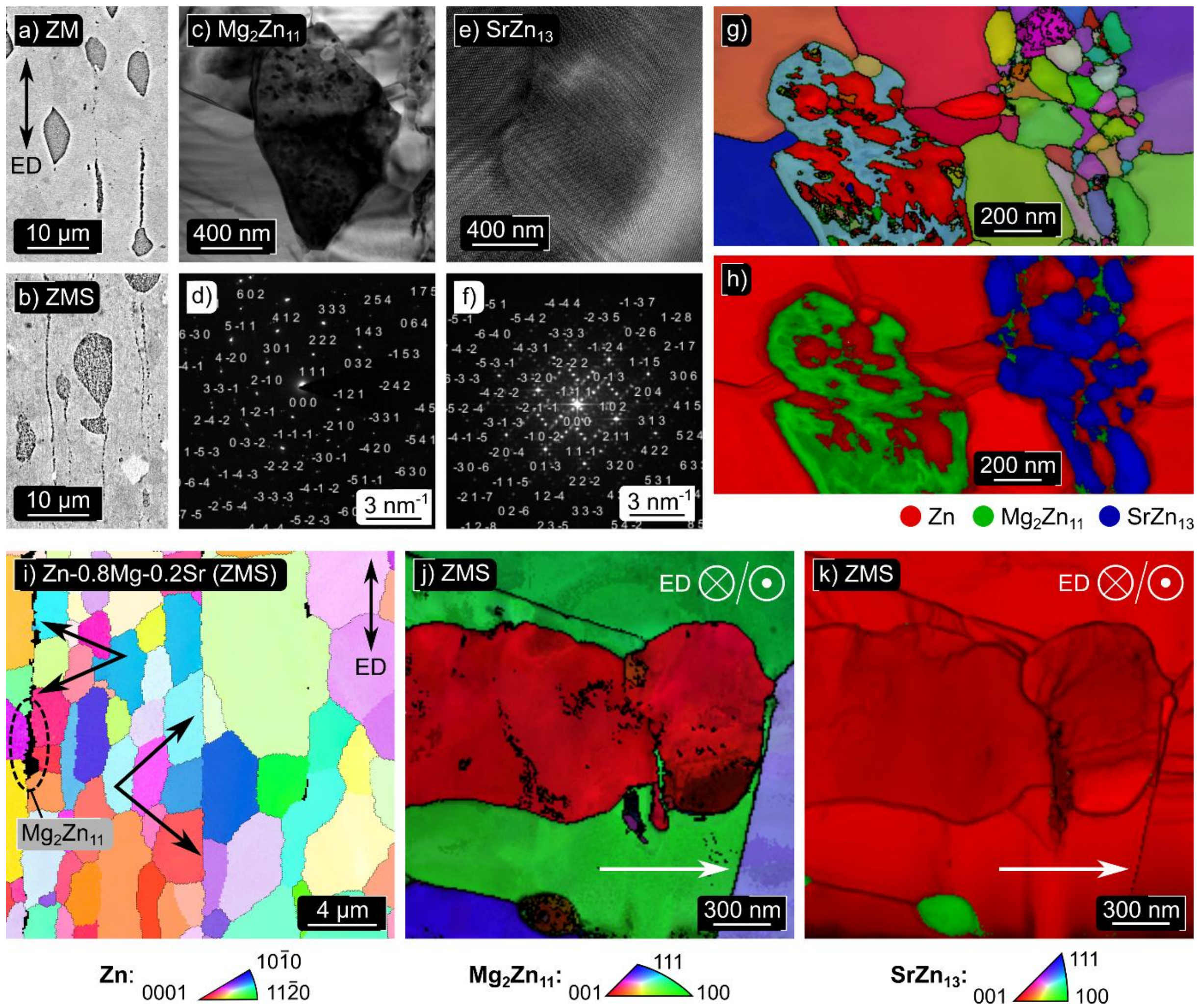

Fig. 3. a), b) BSE SEM images of the phases in alloyed wires. c), d) BF TEM of $Mg_2Zn_{11}$ phase, together with the SAED pattern from $Mg_2Zn_{11}$. e), f) HRTEM image of $SrZn_{13}$ phase and the corresponding FFT pattern. g), h) ASTAR orientation map and phase map from Zn-0.8Mg-0.2Sr wire, respectively. i)–k) IPF map, ASTAR orientation map, and phase map of Zn-0.8Mg-0.2Sr wire, respectively.

The sharp ED-aligned interfaces observed in ZM and ZMS (Figs. 1e and f) were further examined by IPF and ASTAR in Figs. 3i–k for ZMS wire. Two types of such interfaces were observed: the first continued the tail-like structures of fragmented $Mg_2Zn_{11}$ particles, whereas the second occurred with no secondary phase present in the vicinity, both spanning large distances along the ED. ASTAR analysis revealed no intermetallic phases within this interface, and EDS line scans across it showed no associated compositional change; in most cases, the interface corresponds to high-angle grain boundaries between adjacent grains. No similar interfaces have been reported in the literature. Their origin in the alloyed wires is likely associated with fragmentation of $Mg_2Zn_{11}$ phases, as a much higher interface density was observed in ZMS compared to ZM. Moreover, such interfaces likely contribute to a higher number of DRX-deformed fraction in ZMS compared to ZM (19 vs. 9 %, respectively) by accommodating local strain as rigid-body-like sliding along the interface, with larger GOS values and GS observed between such interfaces, despite the finer overall GS observed in ZMS.

## 3.2 Tensile testing

Representative engineering stress-strain curves for PZ, ZM and ZMS wires tested at RT of around 24 °C, at 37 °C, and at RT after 20 days of aging at 37 °C are shown in Fig. 4a, with magnified regions in Figs. 4b–d. For each alloy, the three testing conditions result in curves with comparable overall shape, differing only slightly in stress level or in strain to fracture, except for ZM after 20 days of aging. PZ showed the lowest stress of the three compositions, with a pronounced serrated flow superimposed on the work-hardening trend throughout the plastic region for all conditions, disappearing beyond roughly 10 % elongation (Figs. 4a and b). ZM at RT and 37 °C showed a single, pronounced stress drop shortly after the ultimate tensile strength, followed by partial recovery and renewed softening toward fracture at comparatively low strain (Fig. 4c). This drop became progressively less pronounced with aging, nearly disappearing after 10 and 20 days at 37 °C. ZMS, in contrast, showed a smaller, qualitatively different stress drop just after yielding, followed by an extended near-plateau region before gradual softening to fracture (Figs. 4a, d). Across all three alloys, curves from the different testing conditions overlapped closely up to fracture, with the fracture strain primarily varying between conditions, generally within the scatter shown in Fig. 4g.

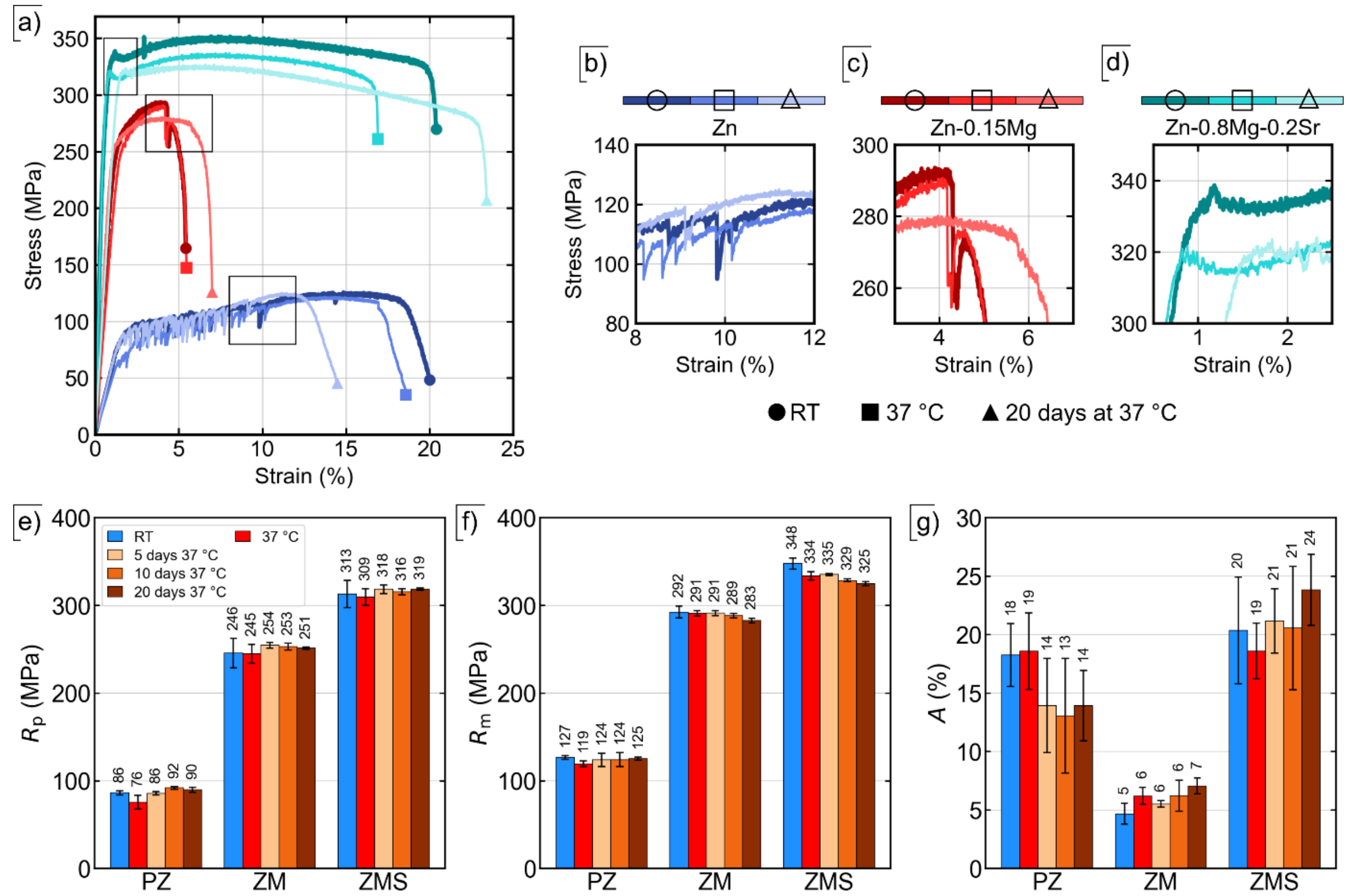

Fig. 4. a) Representative tensile test curves measured at RT, 37 °C, and at RT after 20 days aging at 37 °C, with insets shown in b)–d). e)–g) The 0.2 % offset yield strength, ultimate tensile strength, and elongation to failure for different conditions (for numerical values, see Table S2, Supplementary Materials).

$R_p$ and $R_m$ varied only slightly across the five testing conditions for a given alloy, generally within the error bars (Figs. 4e and f). $R_p$ increased from 76–92 MPa for PZ to 245–254 MPa for ZM and 309–319 MPa for ZMS, and $R_m$ from 119–127 MPa for PZ to 283–292 MPa for ZM and 325–348 MPa for ZMS, depending on the condition. For the alloyed wires, the scatter in $R_p$ and $R_m$ decreased when testing at 37 °C and with prolonged aging, possibly reflecting mild diffusion-driven homogenization at this temperature. $A$ did not follow the same simple trend with alloying and showed markedly larger scatter across conditions (Fig. 4g). PZ reached a slightly lower $A$ of 13–14 % after aging than at RT or 37 °C (18–19 %). ZM showed the lowest

*A* of the three alloys at every condition (5–7 %) but with the smallest scatter, while ZMS showed the highest *A* (19–24 %) but also the largest scatter.

The marked increase in $R_p$ and $R_m$ with alloying is consistent with the microstructural changes described in Section 3.1. Both ZM and ZMS show substantially finer GS than PZ, contributing to grain-boundary strengthening, together with a large fraction of the harder $Mg_2Zn_{11}$ (and, for ZMS, $SrZn_{13}$) phase, which can additionally strengthen the wires through load transfer and impediment of dislocation motion [13]. The higher $R_p$ and $R_m$ of ZMS relative to ZM are consistent with its finer GS and higher total secondary-phase content, including the harder $SrZn_{13}$ phase, while the effect of the phases on elongation is addressed in Section 4.3. As digital image correlation (DIC) could not be reliably applied to specimens of this diameter, Young's modulus is not reported. However, the slope of the elastic region of the tensile curves (Fig. 4a) suggests an increasing elastic modulus from PZ to ZM and to ZMS.

Fracture surfaces of tensile specimens tested at 37 °C are shown in Fig. 5. Low-magnification overviews (Figs. 5a–c) include a circle denoting the AE wire diameter of ~290 µm. PZ fractured by transgranular cleavage with tearing ridges and fine river markings, without significant signs of ductile behavior. Alloying with Mg refined the microstructure and correspondingly roughened the fracture surface. ZM showed cleavage facets with only limited ductility and river markings (Fig. 5e) similar to PZ, while ZMS showed visibly higher roughness, attributed to its finer grain size and cleavage facets extending through the intermetallic regions surrounded by elongated, ductile Zn matrix edges. Both ZM and ZMS additionally showed particles with a cleavage-like fracture surface (Figs. 5j and k), indicating brittle failure of the particles during tensile loading. Further analysis of the necking region via µCT is discussed in Section 4 in the Supplementary Materials (Fig. S5).

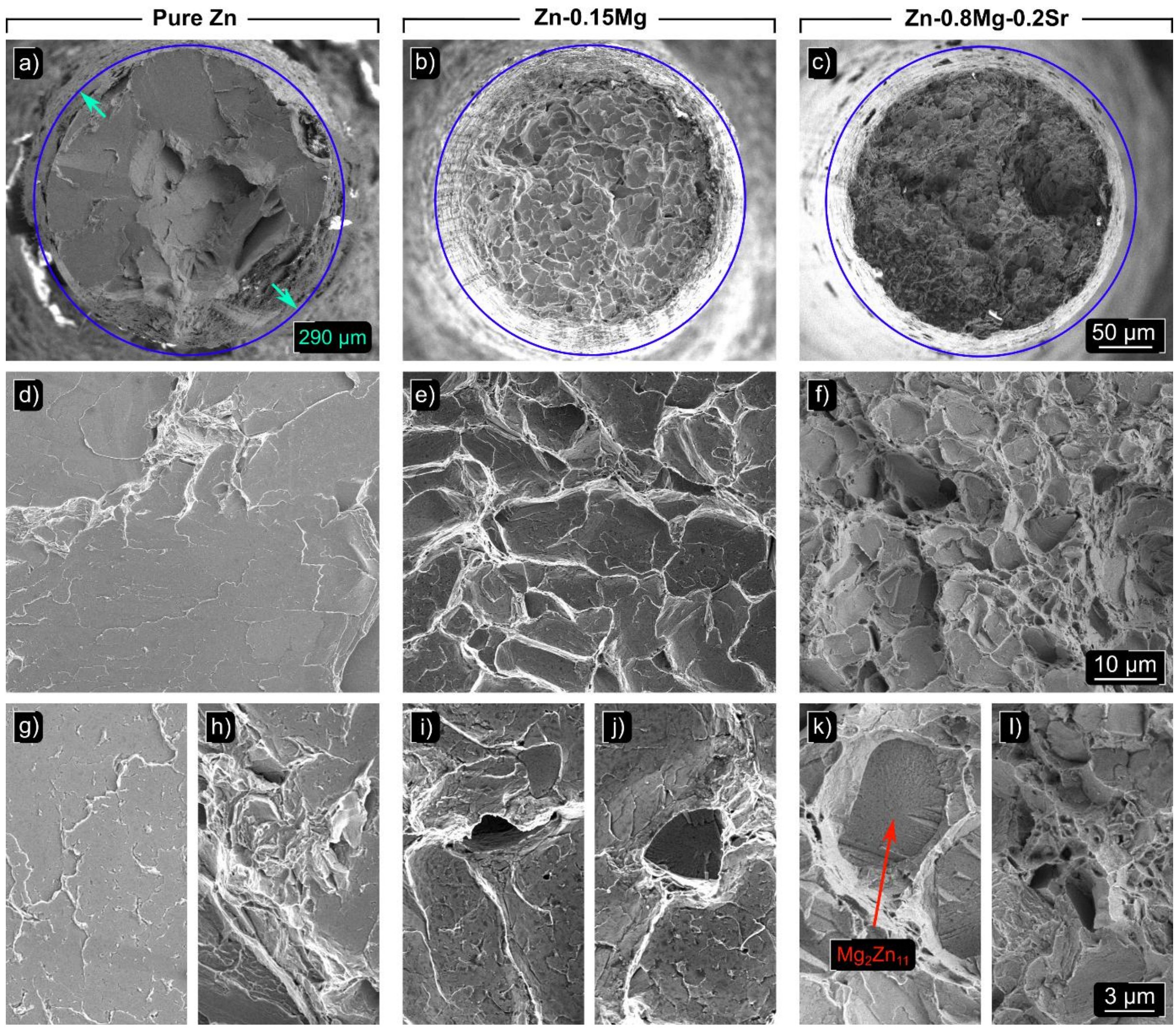

Fig. 5. SE SEM fracture surface analysis after tensile test at 37 °C, with images for a given row sharing the magnification bar.

Microstructure analysis of the neck regions formed after tensile testing at 37 °C is shown in Fig. 6, with corresponding EBSD maps in Fig. S6 (Supplementary Materials). Equivalent analysis on RT-tested specimens showed no measurable difference from the 37 °C specimens. SEM BSE images (Figs. 6b and c) show fractured $Mg_2Zn_{11}$ particles in ZM and ZMS, consistent with the particle cracking observed by fractography (Fig. 5). Crystal direction maps (CD) near the fracture surface in Figs. 6d–f show extensive compression twin (CT) formation in ZM as green color, markedly fewer CTs in ZMS, and only a limited amount of CTs in PZ. CT formation is consistent with the basal-fiber texture inherited from extrusion (Section 3.1), where the c-axis of most grains lies close to perpendicular to the ED. This orientation favors compression twinning, while the Schmid factor for basal slip under axial tension is very low, so the majority basal-textured grain population is expected to accommodate tensile straining preferentially by twinning. In reality, this reasoning is simplified, since the fiber texture is only statistically defined and the stress state within the cylindrical wire cross-section deviates from ideal uniaxial loading, yet it still largely accounts for the observed compression twinning. The GOS map at the neck for PZ (Fig. 6g) shows a largely recrystallized and finer-grained microstructure compared to the AE state (Fig. 1j). In contrast, ZM and ZMS show much higher GOS values at the neck (Figs. 6h and i) than in the AE microstructure (Figs. 1k and l), reflecting the plastic strain imposed by tensile loading and necking. For all wire compositions, this strain is sufficient to trigger DRX near the fracture surface, seen as low-GOS grains interspersed among deformed regions, with the number of DRXed grains being the highest in PZ and lowest in ZM.

GOS maps acquired approximately 350 μm from the neck (Figs. 6j–l) revealed markedly different behavior between the alloys. PZ remained predominantly high-GOS with fewer recrystallized grains compared to the neck. For ZM, the GOS distribution at this distance closely resembled the AE state, indicating that the additional straining from necking remained sharply confined to its vicinity. ZMS, however, showed a GOS distribution at 350 μm already substantially higher than its AE state and close to that observed at its own neck, though with a much lower recrystallized fraction. This suggests that, unlike ZM, plastic straining in ZMS is accommodated over a considerably longer distance before final necking and fracture.

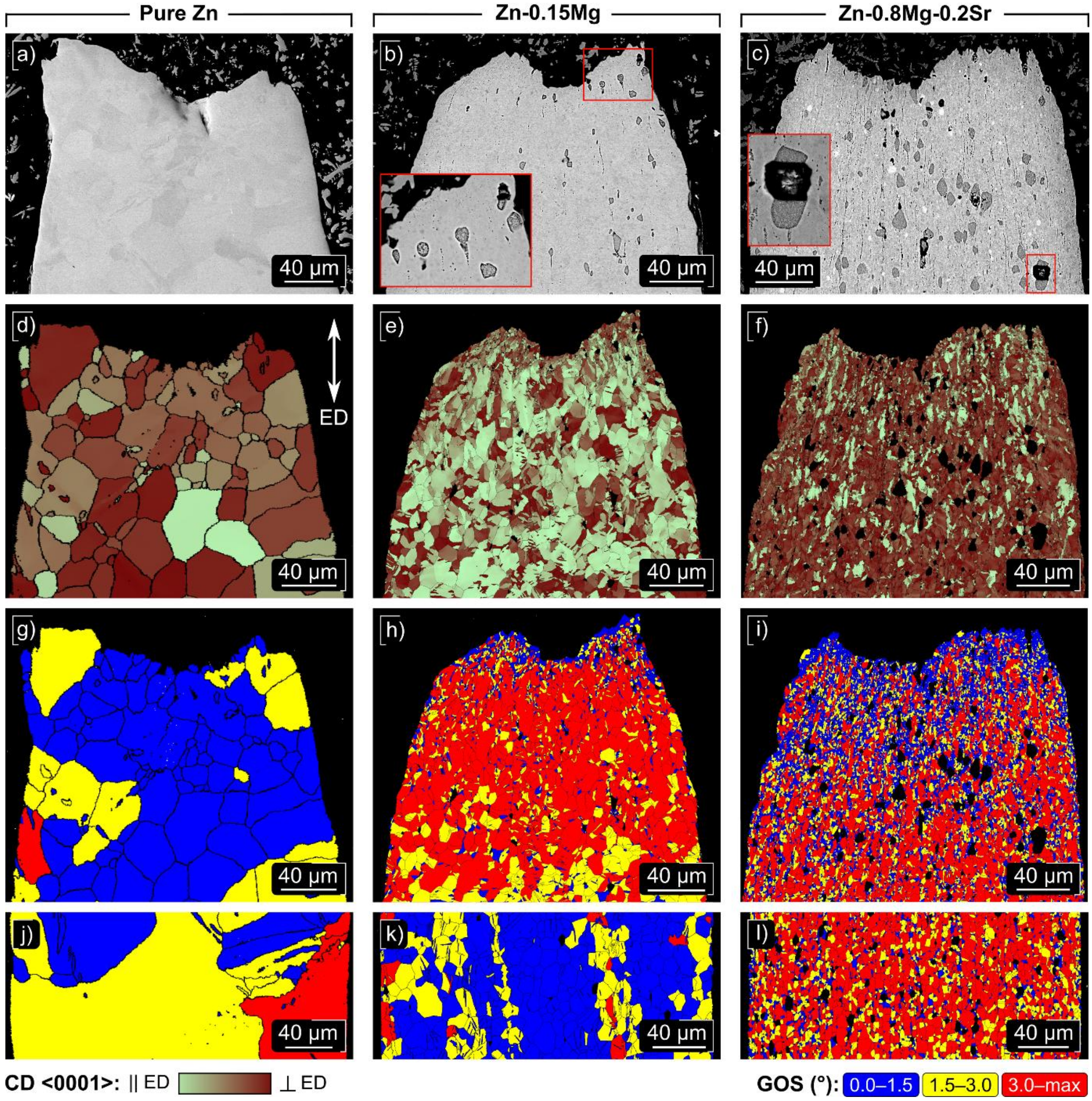

Fig. 6. a)–c) BSE SEM images of the neck regions following tensile testing at 37 °C. d)–f) Corresponding CD maps, together with GOS maps in g)–i). j)–l) GOS maps collected approximately 350 µm from the neck region.

### 3.3 Bending behavior

Qualitative bending behavior of PZ, ZM and ZMS wires, assessed by EBSD, is shown in Fig. 7, with each row corresponding to a bending condition of: one bending (B1), bending and straightening (S1), and B1 with bending to the opposite direction (B2). For each wire and condition, GOS maps were used to assess the extent of plastic deformation, while CD maps show twin formation. After one bending (B1), PZ showed only a slight GOS increase relative to the AE state (Fig. 7a), with extensive compression twinning in the tensile zone (TZ) and none in the compression zone (CZ), together with grain growth compared to the AE state. ZM showed a larger GOS increase in both zones (Fig. 7b), with the neutral zone shifted toward the CZ and extensive CT formation confined to the TZ, similar to PZ. ZMS showed qualitatively similar GOS behavior to ZM, with the neutral zone shifted toward the CZ, but lower GOS values in both zones and only very limited CT formation in the TZ (Fig. 7c).

Straightening (S1) reverses the CZ and TZ of the bent wire. For PZ, S1 produced new CTs in the former CZ (now TZ) together with secondary twinning of the original CTs in the former TZ (now CZ), recovering the AE c-axis orientation (Fig. 7d). This twinning-detwinning mechanism is well documented for magnesium and Mg-based alloys [41,42] but, to our knowledge, has not been previously reported for zinc or its alloys. Further bending to B2 increased GOS values, with a complete twinning-detwinning mechanism in the new CZ and twinning in the TZ (Fig. 7g).

For ZM, S1 gave GOS values similar to B1 in both zones, with some remaining twins in the previous TZ. Further bending in B2 showed full twinning-detwinning in the CZ and slightly more CT formation in the TZ than after B1 (Figs. 7e and h). ZMS behaved similarly during S1, with GOS increasing in the previously neutral zone and CTs forming in the TZ with some remaining in the CZ. B2 once again produced a neutral zone, twinning-detwinning in the CZ, and more extensive TZ twinning than after B1, though the neutral-zone shift during B1 was less pronounced than for ZM, and GOS values in both zones remained lower than for ZM throughout (Figs. 7f, i). ZM and ZMS therefore showed qualitatively similar bending behavior overall, differing mainly in the extent of twin formation and GOS values in the TZ and CZ.

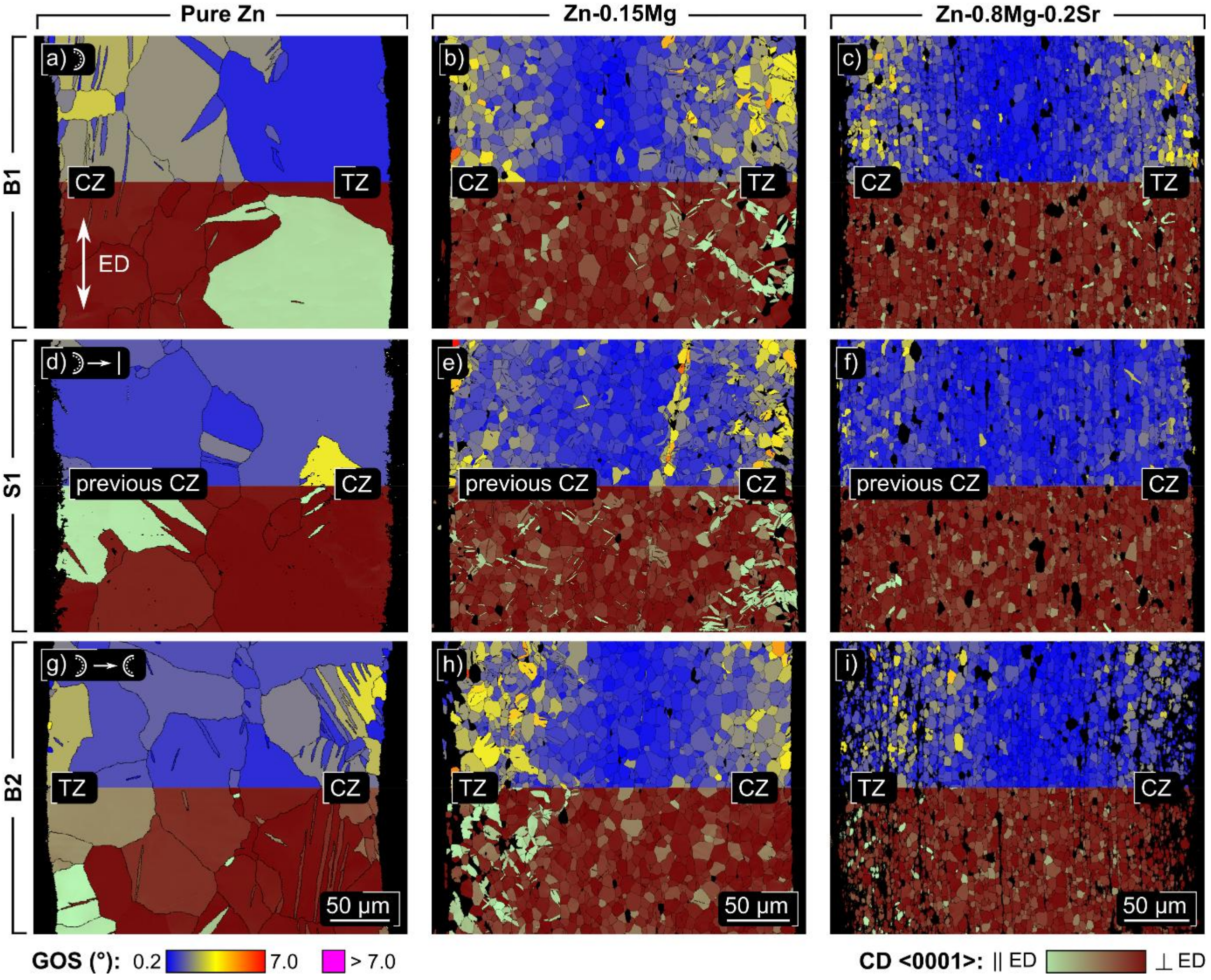

Fig. 7. GOS and CD maps after: a)–c) one bending, d)–f) one bending and straightening, g)–i) one bending with bending to the opposite direction.

## 4 Discussion

### 4.1 Microstructure evolution during extrusion and necking

Alloying weakens the basal-fiber texture from 17.1 m.r.d. in PZ to 4.4 and 3.9 m.r.d. in ZM and ZMS, consistent with $Mg_2Zn_{11}$ and $SrZn_{13}$ promoting particle-stimulated nucleation (PSN) of new, randomly oriented grains during DRX [43]. This effect is well reported for other alloyed hcp systems as well [44,45], and is consistent with the predominantly low-GOS microstructure of ZM and ZMS, compared to PZ (Figs. 1j–l).

A more detailed PF analysis in Figs. 8a–e showed a near-surface texture intensification for both alloyed wires with m.r.d. rising from 6.3/5.1 in the middle of the wire to 14.5/11.1 at the edges for ZM/ZMS, consistent with friction-induced shear at the die wall superimposed on axisymmetric compression governing deformation in the middle [46] and reported for directly extruded Mg-Zn wires in our previous study [42]. Furthermore, μCT revealed a radial dependence of intermetallic particle content, depleted near the surface and toward the middle, with a maximum closer to the surface. This distribution plausibly reflects shear-induced particle migration away from the high-shear near-surface region and the resulting stress gradient towards the wire axis [47] combined with a limited μCT detection resolution for the finest particles.

The markedly greater twin extent in ZM than ZMS at the neck (Figs. 6e, f) reflects both its coarser grain size of 6.8 vs. 3.3 μm in the AE state, given the increase in critical twinning stress with decreasing grain size [50], and higher secondary-phase content in ZMS, which may promote additional slip-based strain accommodation around particles, competing with twin activation. The comparatively modest twin activity in PZ based on the CD map in Fig. 6d, despite its far coarser grain size, is explained by DRX progressively consuming the original twinned regions as new grains nucleate from the twin substructure. The larger number of recrystallized grains at the ZMS neck relative to ZM is consistent with ZMS's higher intrinsic susceptibility to PSN-assisted DRX, though the much lower overall elongation of ZM before failure means its neck also accumulates less strain, so a weaker DRX response may simply reflect insufficient strain energy rather than an intrinsic compositional difference.

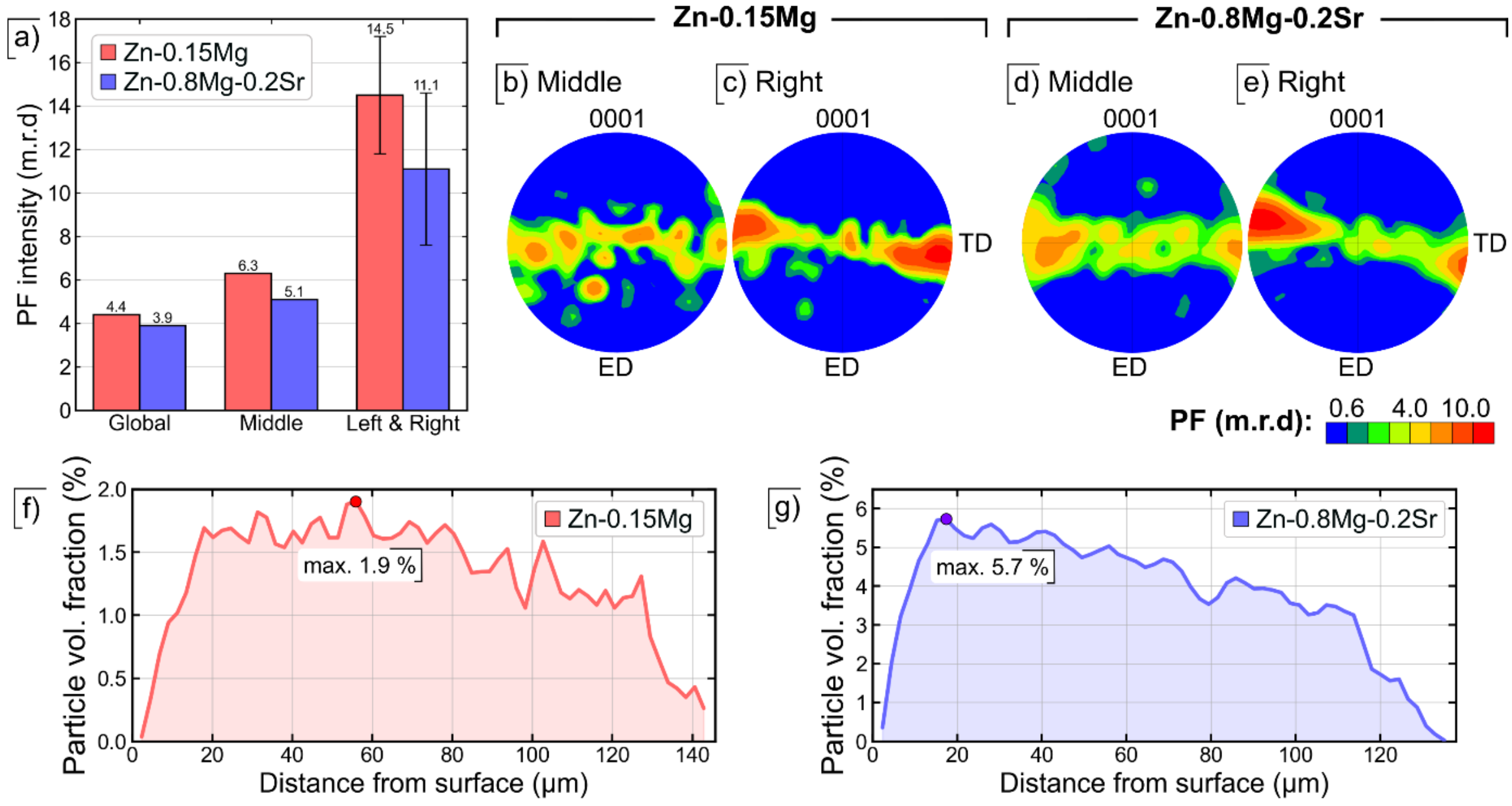

Fig. 8. a) PF intensity (m.r.d.) for the full c ross-section (global), the middle region, and the averaged left and right edge regions of ZM and ZMS wires. b), c) PFs of Zn-0.15Mg wire in the middle and right-edge regions, together with d) and e) analogous analysis for Zn-0.8Mg-0.2Sr wire. f), g) Intermetallic particle volume fraction as a function of distance from the wire surface for ZM and ZMS wires.

## 4.2 Serrated flow and stress drops during tensile deformation

The repeated serrations in the PZ tensile curve (Fig. 4b) reflect deformation-twinning events transmitted across weakly misoriented grains in the basal-fiber texture, each spanning a substantial fraction of the cross-section and producing an abrupt, discrete stress drop [48]. CT bands oriented at ~45 ° to the wire axis in Fig. S2c (Supplementary Materials) align with the plane of maximum resolved shear stress for twinning, with cross-section-spanning twins independently confirmed on the wire surface during DIC attempts. Serrations disappear beyond ~10 % elongation because twinning is polar and orientation-dependent, so the most favorably oriented grains twin first and further strain is instead accommodated by the harder non-basal slip. The PZ fracture surface nonetheless retains a quasi-cleavage character despite the ductile curve, suggesting final failure nucleates locally (e.g. at a twin-matrix interface) rather than reflecting the bulk deformation mode. A similarly serrated response was reported for coarse-grained extruded pure Zn wires [49], though those wires fractured directly within the serrated regime rather than transitioning to smooth flow, plausibly reflecting their absence of a basal fiber texture. The substantially finer grain size of ZM and ZMS suppresses twinning [50], explaining the absence of repeated serrations. Instead, ZM shows a single, pronounced stress drop near the ultimate tensile stress, associated with strain localization and extensive twinning at the neck (Figs. 9d and e) and more plausibly reflecting a discrete damage event, either a fracture of a large $Mg_2Zn_{11}$ particle or coalescence of cracked particles, rather than a twinning-related origin. Supporting ex-situ μCT identified pores at fractured $Mg_2Zn_{11}$ particles confined to the neck region (Figs. 9a–c), with a comparable porosity increase toward the fracture surface also observed for ZMS (Fig. 9f). Because ZM contains fewer particles than ZMS, a crack nucleated at a fractured particle propagates over a larger, less impeded distance, favoring one resolvable drop, whereas ZMS's much higher particle density distributes cracking smoothly over the deformation history without a comparable dip.

The stress drop seen for ZMS shortly after yielding (Fig. 4d) reflects a classical yield-point phenomenon linked to scarce mobile dislocations in fully recrystallized material [51], confirmed by its absence upon loading beyond $R_p$, unloading, and subsequent loading (see Fig. S7, Supplementary Materials), and was not observed for every ZMS specimen, plausibly reflecting specimen-to-specimen microstructural variability.

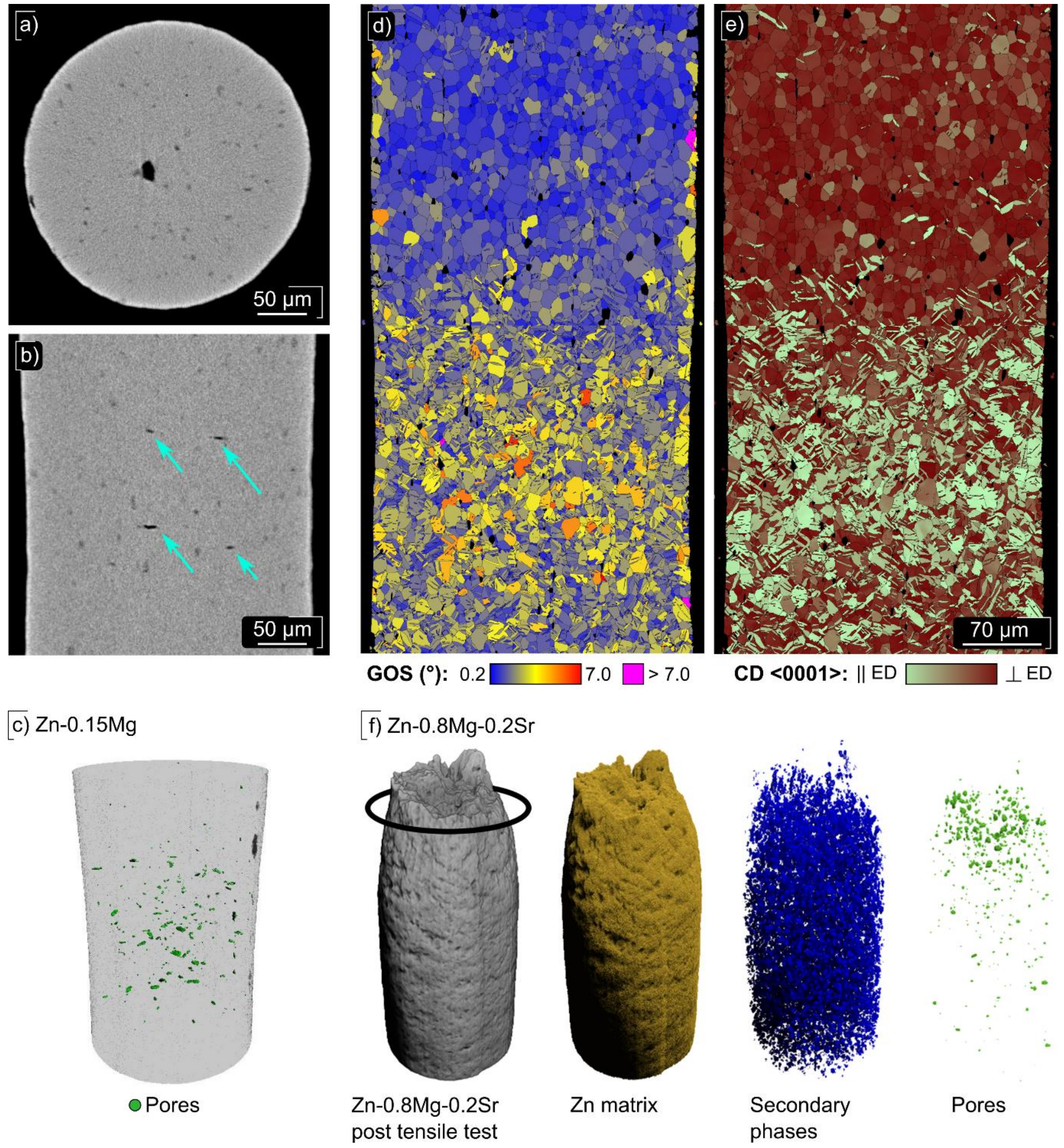

Fig. 9. a)–c) µCT analysis of Zn-0.15Mg wire prior to rupture (stopped manually after the dip observed in Fig. 4c), together with GOS and CD maps in d) and e). f) µCT analysis of the Zn-0.8Mg-0.2Sr wire after tensile test. The black ring shows the diameter of an AE wire.

### 4.3 Ductility differences and the influence of aging

ZM and ZMS differ markedly in mean elongation to failure (6 vs. 19 %). A contributing factor might be the more elongated, ED-aligned $Mg_2Zn_{11}$ morphology in ZMS, which orients particles favorably for damage resistance under tensile loading [52]. However, the primary contributor is likely the far more extensive neck localization and twinning in ZM (Fig. 6 and Fig. 9), since extensive twinning in hcp metals promotes crack initiation via twin-twin and twin-dislocation interactions [34], previously linked to reducing *A* in coarse-grained, twinning-dominated Zn [49]. This twinning and GOS elevation in ZM remains sharply confined to a small volume around the neck, plausibly reflecting a locally self-reinforcing twinning process that progressively accumulates crack-initiation sites within a small volume. The ZM stress drops progressively weakened with aging at 37 °C, and nearly disappeared after 10 and 20 days of aging, suggesting a thermally activated relaxation of residual stress and dislocation density at the matrix-precipitate interface, yet *A* increased only marginally, confirming ductility is not strictly governed by this discrete damage event.

Elongation scatter also differed between compositions (Fig. 4), with ZM showing the smallest scatter alongside its lowest *A*, since fracture follows shortly after neck localization, while ZMS showed comparatively larger scatter, most pronounced after aging, reflecting its much larger total plastic strain and correspondingly more opportunity for local differences to compound. Larger scatters in PZ likely reflect its rougher AE surface and quasi-cleavage-dominated fracture (Fig. S3, Supplementary Materials, and Fig. 5, respectively).

### 4.4 Bending behavior of the wires

To our knowledge, the twinning-detwinning mechanism observed here by EBSD in Fig. 7 has not been previously reported experimentally for Zn or its alloys, though it is well established for Mg [41]. For Zn specifically, the reversal response during loading has been addressed theoretically via a crystal-plasticity model of forward-reverse shear in rolled Zn-Cu-Ti sheet, but finding twinning's contribution negligible and attributing the response instead to Bauschinger-effect dislocation mechanisms [53].

Twinning-detwinning (twinning in AE-basal texture, with secondary twinning recovering the orientation similar to the AE state) follows directly from the crystallographic polarity of $\{10\bar{1}2\}$ compression twinning in Zn. During the first bending (B1), the tension zone (TZ) is crystallographically equivalent to c-axis compression via strain, favoring CT formation, while the compression zone (CZ) corresponds to c-axis extension, for which Zn has no complementary twin, so strain is instead accommodated by non-basal slip. Twinning is consequently solely confined to the TZ, consistent with CT formation in the literature for pure Zn [11] and with the observed CTs during tensile testing in Fig. 6. Upon straightening (S1), the zones reverse. The previous CZ with no CTs nucleates twins due to similar c-axis strain compression, while the previously twinned region (now CZ) develops a secondary set of twins resulting from compressive stress, rather than strain, in the c-axis that restores an orientation close to the basal-textured parent grain. An analogous cycle was reported for basal-textured Mg-Zn wires, but with TZ and CZ inverted, since $\{10\bar{1}2\}$ twinning is an extension rather than compression twin in Mg [42] and is therefore favored in opposite deformation zones.

For both alloyed wires, the neutral zone shifted from the geometric center toward the CZ, consistent with the known tension-compression yield asymmetry of basal-textured Zn (CYS/TYS ≈ 1.93, TYS 113.8 ± 1.6 and CYS 219.5 ± 1.8 MPa [10]). Since the CZ sustains a higher stress before yielding, the neutral axis must shift toward it so that the resulting larger, lower-stress TZ balances the smaller, higher-stress CZ and the net axial force across the cross-section remains zero.

The bending experiments revealed the most significant limitation of the present wires. Cyclic bending produced cracking and fracture within only 3–6 cycles for pure zinc, and repeatedly during the third cycle for both alloyed wires, despite their vastly different secondary-phase content (1.3 and 4.8 vol%). The ZMS wire in Figs. 10a–e with a bending-induced crack was imaged prior to the development of the bending apparatus used for Fig. 7, precluding a direct comparison with those results. Nevertheless, a substantial grain refinement was observed at the wire edges, consistent with continuous DRX (CDRX) driven by dislocations reintroduced during cyclic bending [54]. Similar cracking after cyclic bending for ZM is depicted in Figs. 10f and g.

Cracks for both alloyed wires consistently initiated at the wire surface rather than at $Mg_2Zn_{11}$ particles, which indicates that the limiting factor is not particle-driven but reflects the intrinsically limited number of independent, low-stress deformation modes in Zn [8,9]. Furthermore, each bending reversal requires the twinning-detwinning cycle to operate, regenerating twin-twin interfaces that are well-documented crack-nucleation sites in hcp metals [34]. Shifting the extrusion texture away from the strong basal fiber component, e.g., toward orientations favoring basal or pyramidal slip over the polar twinning mode [35] is a plausible route to improve cyclic bendability, though texture broadening must be approached carefully since it can instead promote microcracking at highly misoriented boundaries [55].

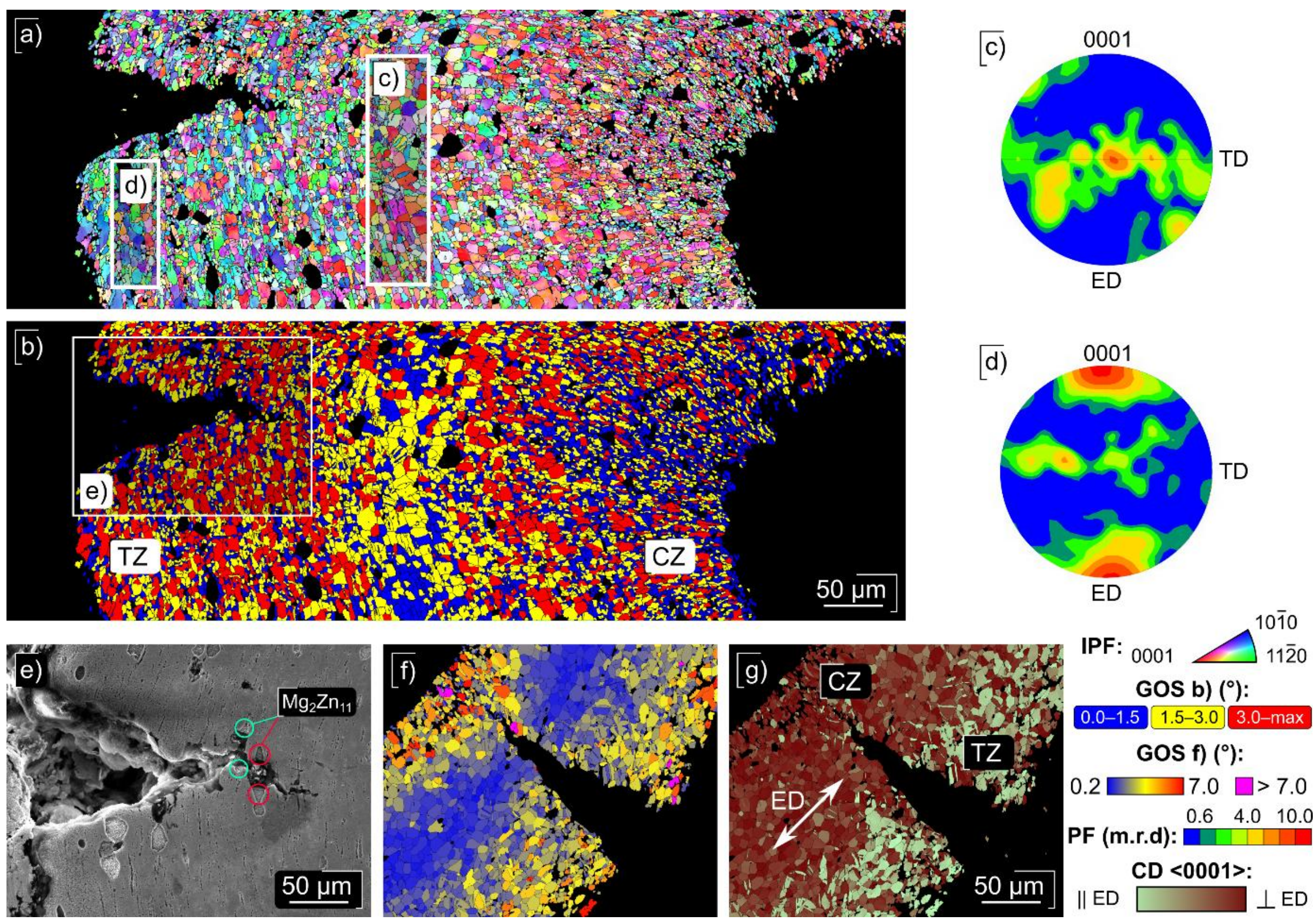

Fig. 10. a), b) IPF and GOS map of Zn-0.8Mg-0.2Sr wire after cyclic bending, together with PF in c) and d) from different regions of the deformed wire. e) SE SEM detail of the observed crack. f) and g) GOS and CD map of a fractured Zn-0.15 wire after cyclic bending.

## 5 Conclusion

In this study, Zn, Zn–0.15Mg (ZM), and Zn–0.8Mg–0.2Sr (ZMS) thin wires were prepared by one-step direct extrusion with an extrusion ratio of ~1:400 and subjected to detailed microstructural and mechanical characterization. The most significant observations can be summarized as follows:

- The presence of secondary phases ($Mg_2Zn_{11}$ and $SrZn_{13}$) led to microstructural refinement through the PSN mechanism, accompanied by a weakening of the basal texture compared with pure Zn.
- µCT and SAXS revealed a multimodal distribution of intermetallic particles, with a simultaneous concentration gradient toward the wire surface and centre.
- The tensile yield strength at 37 °C increased from approximately 76 MPa for PZ to 245 MPa for ZM and further to 309 MPa for ZMS, with only minor changes after aging.
- All compositions exhibited markedly poor cyclic bendability, governed by the limited number of deformation modes available in basal-textured Zn. A twinning–detwinning mechanism at the grain level was directly observed for the first time in Zn.

In conclusion, alloying with Mg and Mg+Sr provides an effective route for refining and strengthening extruded Zn wires while maintaining their short-term stability under thermal exposure. However, poor cyclic bendability remains an intrinsic, composition-independent limitation. Addressing this limitation is therefore a key direction for further development of these wires for bending-critical biomedical applications, such as guidewires and cerclage wires.

**CRediT authorship contribution statement**

**L. Hlodák**: Conceptualization, Methodology, Formal analysis, Validation, Investigation, Data curation, Writing – original draft, Writing – review & editing, Visualization. **K. Tesař**: Conceptualization, Methodology, Formal analysis, Validation, Investigation, Resources, Writing – review & editing, Visualization, Supervision, Project administration, Funding acquisition**. M. Lebeda**: Formal analysis, Investigation, Writing – original draft. **J. Duchoň**: Formal analysis, Investigation. **J. Kubásek**: Investigation. **J. Čech**: Methodology, Validation, Formal analysis, Investigation. **A. Školáková**: Investigation. **J. Pinc**: Conceptualization, Methodology, Formal analysis, Investigation, Resources, Writing – original draft, Writing – review & editing, Visualization, Supervision.

**Acknowledgments**

The authors acknowledge the support provided by the project Resorbable metal implants for reconstructive hand surgery, supported by the Ministry of the Environment of the Czech Republic; project No. CZ.10.02.01/00/25_082/0001369, co-funded by the European Union. L.H., K.T., J.Č. and J.P. acknowledge the support provided by the Ferroic Multifunctionalities project, supported by the Ministry of Education, Youth, and Sports of the Czech Republic; Project No. CZ.02.01.01/00/22_008/0004591, co-funded by the European Union. L.H. and K.T. were further supported by the Czech Science Foundation (GAČR), Junior Star project No. 25-17788M. J.K. and J.P. were supported by the Czech Science Foundation (GAČR) project No. 25-16144S; J.D., A. Š. were supported by the CzechNanoLab project LM2023051, funded by the Ministry of Education, Youth and Sports of the Czech Republic, which is gratefully acknowledged for enabling measurements and sample fabrication at the LNSM Research Infrastructure.

We sincerely thank Jiří Ryjáček, Ondrej Kovářík, Štěpán Hortlík, Jan Drahokoupil, and Petr Svora for their technical assistance and discussions.

**Data availability**

Data are available online on Zenodo (DOI: https://doi.org/10.5281/zenodo.22856727) and on request.

**Declaration of generative AI and AI-assisted technologies in the manuscript preparation process**

During the preparation of this work the author(s) used Claude by Anthropic in order to assist with language polishing, grammar correction, and improvement of textual clarity and flow. After using this tool/service, the author(s) reviewed and edited the content as needed and take(s) full responsibility for the content of the published article.